\documentclass[prl, twocolumn, preprintnumbers, nofootinbib]{revtex4}
\pdfoutput=1

\usepackage[utf8]{inputenc}
\usepackage{amsmath, amssymb}
\usepackage{graphicx, xcolor}

\newcommand{\rR}{\rho_R}
\newcommand{\rbh}{\rho_\text{BH}}
\newcommand{\gs}{g_\star}
\newcommand{\gss}{g_{\star s}}
\newcommand{\Trh}{T_\text{rh}}
\newcommand{\Tmax}{T_\text{max}}
\newcommand{\Min}{M_\text{in}}
\newcommand{\Tin}{T_\text{in}}
\newcommand{\mdm}{m_\text{DM}}
\newcommand{\Ndm}{N_\text{DM}}
\newcommand{\gdm}{g_\text{DM}}
\newcommand{\Cn}{\mathcal{C}_\text{DM}}
\newcommand{\Tbhin}{T_\text{BH}^\text{in}}

\begin{document}

\title{Gravitational dark matter production:\\primordial black holes and UV freeze-in.}

\author{Nicolás Bernal}
\email{nicolas.bernal@uan.edu.co}
\affiliation{Centro de Investigaciones, Universidad Antonio Nariño, Carrera 3 Este \# 47A-15, Bogotá, Colombia.}
\author{Óscar Zapata}
\email{oalberto.zapata@udea.edu.co}
\affiliation{Instituto de Física, Universidad de Antioquia, Calle 70 \# 52-21, Apartado Aéreo 1226, Medellín, Colombia}
\affiliation{Abdus Salam International Centre for Theoretical Physics, Strada Costiera 11, 34151, Trieste, Italy.}

\begin{abstract}
    Dark matter (DM) interacting only gravitationally with the standard model could have been produced in the early universe by Hawking evaporation of primordial black holes (PBH).
    This mechanism is viable in a large range of DM mass, spanning up to the Planck scale.
    However, DM is also unavoidably produced by the irreducible UV gravitational freeze-in.
    We show that the latter mechanism sets strong bounds, excluding large regions of the parameter space favored by PBH production.
\end{abstract}
\preprint{PI/UAN-2020-682FT}
\maketitle

\section{Introduction}

Dark matter (DM) reaching chemical equilibrium with the standard model (SM) can be produced in the early universe via the WIMP paradigm~\cite{Arcadi:2017kky}.
Alternatively, if the interaction rates between the dark and visible sectors are not strong enough, DM can originate non-thermally through the FIMP mechanism~\cite{McDonald:2001vt, Choi:2005vq, Hall:2009bx, Elahi:2014fsa, Bernal:2017kxu}.

However, it is also conceivable that DM only interacts gravitationally with the SM.
In this scenario, DM can be radiated by primordial black holes (PBH)~\cite{Green:1999yh, Khlopov:2004tn, Dai:2009hx, Fujita:2014hha, Allahverdi:2017sks, Lennon:2017tqq, Morrison:2018xla, Hooper:2019gtx, Chaudhuri:2020wjo, Masina:2020xhk, Baldes:2020nuv, Gondolo:2020uqv, Bernal:2020kse, Bernal:2020bjf, Auffinger:2020afu} during their Hawking evaporation~\cite{Hawking:1974sw}.
This mechanism is viable in a large range of DM mass, spanning up to the Planck scale.
PBHs evaporating before BBN are poorly constrained~\cite{Carr:2009jm, Carr:2020gox} and therefore potentially produce the whole observed DM abundance.

Another purely gravitational DM production mechanism corresponds to the UV freeze-in due to annihilation of SM particles via the $s$-channel exchange of gravitons~\cite{Garny:2015sjg, Tang:2017hvq, Garny:2017kha, Bernal:2018qlk}.
The massless SM gravitons mediate between the dark and the visible sectors, producing DM during the heating epoch.
Being a gravitational process, its contribution is Planck suppressed and can be dominant for high reheating temperatures $\Trh$.
Additionally, it only depends on $\Trh$, the DM mass $\mdm$ and its spin.

In this {\it Letter}, the interplay of these two gravitational DM production mechanisms is studied.
In particular, we show that the parameter space favored by the DM production via Hawking radiation is severely constrained when taking into account the irreducible gravitational UV freeze-in.

\section{Dark Matter from\\Primordial Black Holes}
PBHs formed in a radiation-dominated epoch, when the SM plasma has a temperature $\Tin$, have an initial mass $\Min$ given by~\cite{Carr:2009jm, Carr:2020gox, Masina:2020xhk}
\begin{equation}
    \Min = \frac{4\pi}{3}\,\gamma\,\frac{\rR(\Tin)}{H^3(\Tin)}\,,
\end{equation}
where $\gamma \simeq 0.2$, $\rR$ and $H$ are the SM energy density and the Hubble expansion rate, respectively.

PBH evaporation produces all particles, and in particular extra radiation that can modify successful BBN predictions.
To avoid it, we require PBHs to fully evaporate before BBN time, i.e. $T_\text{BBN}\simeq 4$~MeV~\cite{Sarkar:1995dd, Kawasaki:2000en, Hannestad:2004px, DeBernardis:2008zz, deSalas:2015glj}.
This, together with the upper bound on the inflationary scale reported by the Planck collaboration $H_I \leq 2.5 \times 10^{-5}M_P$~\cite{Akrami:2018odb}, bounds the initial PBH mass
\begin{equation}\label{eq:Min}
    0.1~\text{g} \lesssim \Min \lesssim 2\times 10^8~\text{g}\,.
\end{equation}
PBHs that fully evaporate before BBN are typically poorly constrained.
It has been recently pointed out that the production of gravitational waves (GW) induced by large-scale density perturbations underlain by PBHs could lead to a backreaction problem.
However, it could be avoided if the energy contained in GWs never overtakes the one of the background universe~\cite{Papanikolaou:2020qtd}, especially at the BBN era~\cite{Domenech:2020ssp}.

The DM yield produced by Hawking evaporation of PBHs can be estimated by~\cite{Masina:2020xhk, Baldes:2020nuv, Gondolo:2020uqv, Bernal:2020kse, Bernal:2020bjf}
\begin{equation}\label{eq:DMyield}
    Y_\text{DM} = \frac34 \frac{\gs}{\gss}\, \Ndm \times
    \begin{cases}
        \beta\, \frac{\Tin}{\Min} & \text{for}\,\, \beta \ll \beta_c\,, \\[8pt]
        \frac{\bar T_\text{ev}}{\Min} & \text{for}\,\, \beta \gg \beta_c\,,
    \end{cases}
\end{equation}
where $\gs$ and $\gss$ are the number of relativistic degrees of freedom contributing to the SM energy density and entropy, respectively.
The parameter $\beta$ corresponds to the initial PBH energy density normalized to the SM energy density at the time of formation $T = \Tin$:
\begin{equation}
    \beta \equiv \frac{\rbh(\Tin)}{\rR(\Tin)}\,.
\end{equation}
A PBH domination era can be avoided if $\rR \gg \rbh$ at all times, or equivalently if
\begin{equation}
    \beta < \beta_c \equiv \frac{T_\text{ev}}{\Tin}\,,
\end{equation}
where $T_\text{ev}^4 \equiv 9\gs/10240\times M_P^{10}/\Min^6$ and $\bar T_\text{ev} \equiv 2/\sqrt{3}\times T_\text{ev}$, correspond to the SM temperature at the moment when PBHs complete evaporate, for a universe dominated by SM radiation or PBH, respectively.
Additionally, $\Ndm$, given by
\begin{equation}\label{eq:Ndm}
    \Ndm = \frac{15\,\zeta(3)}{\pi^4}\frac{\gdm\, \Cn}{\gs} \times
    \begin{cases}
        \left(\frac{\Min}{M_P}\right)^2 &\text{for}\,\, \mdm \leq \Tbhin,\\[8pt]
        \left(\frac{M_P}{\mdm}\right)^2 &\text{for}\,\, \mdm \geq \Tbhin,
    \end{cases}
\end{equation}
corresponds to the total number of DM particles radiated by a single PBH, with $\gdm$ the number of DM degrees of freedom, and $\Cn=1$ or $3/4$ for bosonic or fermionic DM, respectively.
Furthermore, $\Tbhin \equiv M_P^2/\Min$ is the initial PBH temperature.
$Y_\text{DM}$ is the DM yield at present satisfying $\mdm\,Y_\text{DM} \simeq 4.3 \times 10^{-10}$~GeV to match the observed DM relic abundance $\Omega_\text{DM} h^2\simeq 0.12$~\cite{Aghanim:2018eyx}.

An upper bound on $\Min$ appears if one requires the PBHs to generate the whole observed DM relic abundance.
In the case where PBHs dominate the universe energy density before their decay ($\beta \gg \beta_c$), Eq.~\eqref{eq:DMyield} gives
\begin{equation}\label{eq:under}
    \Min
    \lesssim 
    \begin{cases}
        \frac{{\mathcal C}^2\,M_P\,\mdm^2}{(\mdm\,Y_\text{DM})^2} &\text{for}\quad \mdm\leq\Tbhin\,,\\[8pt]
        \left[\frac{{\mathcal C}^2\,M_P^9}{\mdm^2\,(\mdm\,Y_\text{DM})^2}\right]^{1/5} &\text{for}\quad \mdm\geq\Tbhin\,,
    \end{cases}
\end{equation}
where ${\mathcal C} \equiv \frac{9\sqrt{3}}{16} \left(\frac{5}{2\gs}\right)^{3/4}\frac{\zeta(3)\,\gdm\,\Cn}{\pi^4}$.
Let us note that PBHs with masses violating the bound in Eq.~\eqref{eq:under} are only able to radiate a fraction of the observed DM abundance.
However, the full density can be produced if Eq.~\eqref{eq:under} is satisfied, for a given value of $\beta$.

Moreover, PBHs radiate ultra-relativistic DM~\cite{Baumann:2007yr, Fujita:2014hha, Morrison:2018xla, Bernal:2020kse} that could have a large free-streaming length that suppresses structure formation at small scales.
It is possible to recast the bounds derived for the case of DM thermal relics, coming from the combined data of the CMB and the Lyman-$\alpha$ forest~\cite{Viel:2005qj}.%
\footnote{See, e.g., Refs.~\cite{Lennon:2017tqq, Baldes:2020nuv, Auffinger:2020afu} for alternative analysis that take into account the full DM phase-space distribution.}
Taking its mean velocity to be $v_\text{DM}\lesssim 1.8\times 10^{-8}$ at the moment of the matter-radiation equality~\cite{Masina:2020xhk} for $\mdm \simeq 3.5$~keV~\cite{Irsic:2017ixq}:
\begin{equation}\label{eq:WDMwithout}
    \frac{\mdm}{1~\text{GeV}}\gtrsim 2\times 10^{-3}\left(\frac{\Min}{\text{g}}\right)^{1/2}.
\end{equation}

\begin{figure}
	\centering
	\includegraphics[height=0.39\textwidth]{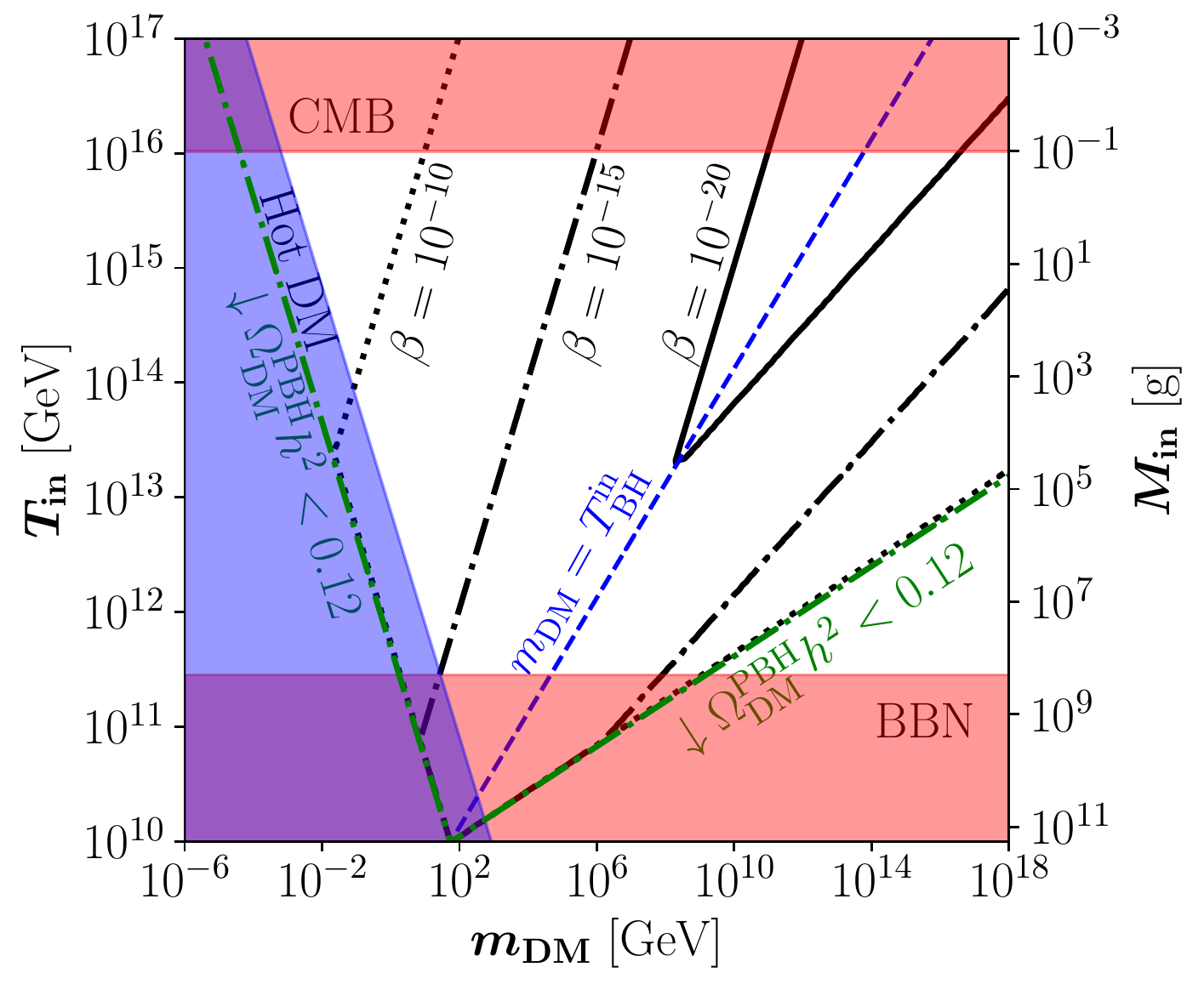}
	\includegraphics[height=0.39\textwidth]{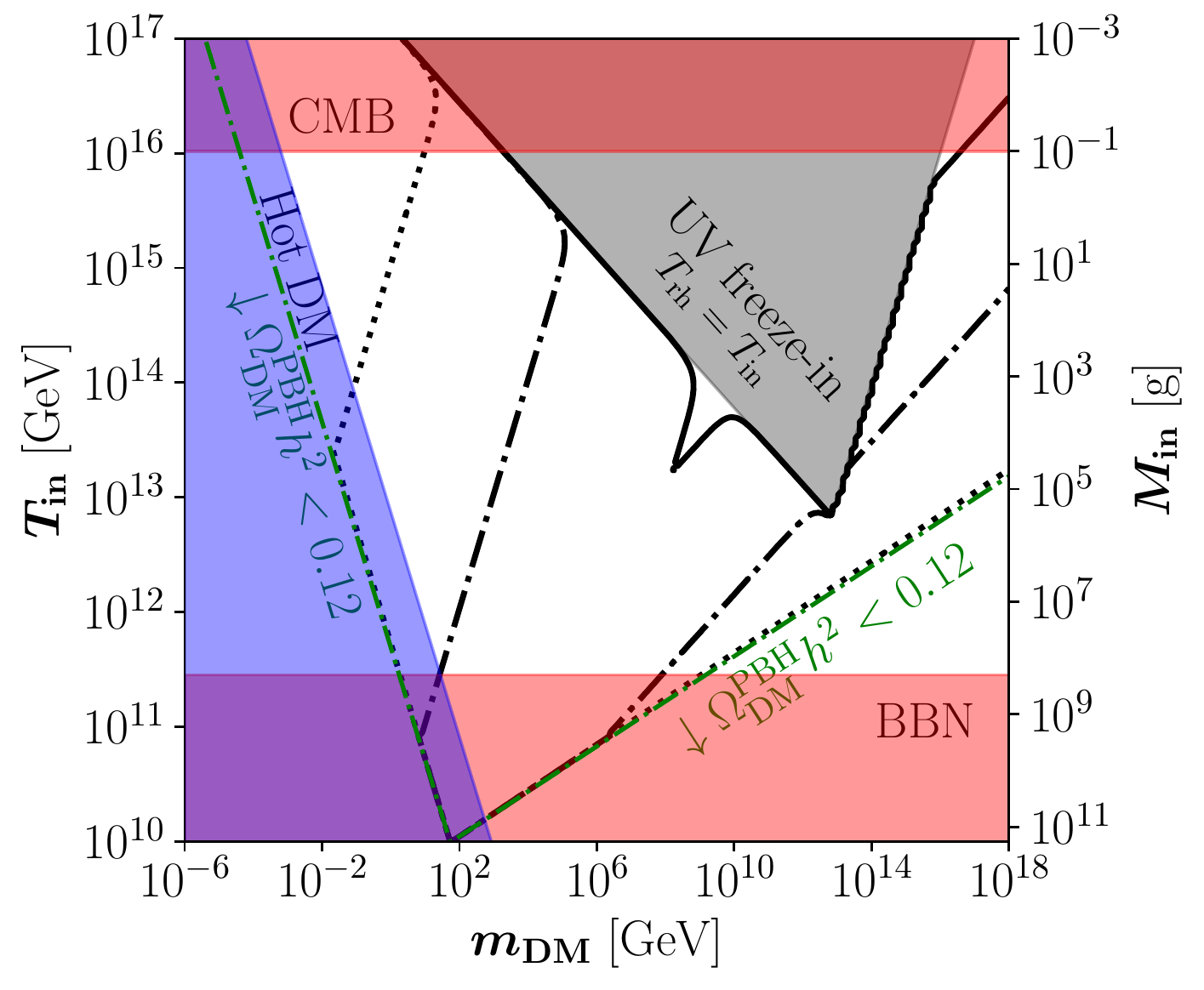}
	\caption{The color shaded areas depict the constraints described in the text.
	Bellow the green dash-dotted lines PBHs produce only a fraction of the total DM.
	The black lines correspond to different values of $\beta$ required to produce the observed DM abundance without (upper panel) and with (lower panel) the irreducible UV freeze-in DM production via the exchange of gravitons, for $\Trh=\Tin$: $\beta=10^{-10}$ (dotted), $\beta=10^{-15}$ (dash-dotted), and $\beta=10^{-20}$ (solid).}
	\label{fig:contour}
\end{figure} 
The color shaded areas in the upper panel of Fig.~\ref{fig:contour} summarize the previously described constraints on the parameter space $[\mdm,\,\Tin]$, for fermionic DM with $\gdm=2$.
Red corresponds to the CMB and BBN bounds in Eq.~\eqref{eq:Min}, and blue to hot DM in Eq.~\eqref{eq:WDMwithout}.
Below the green dash-dotted lines, PBHs produce only a fraction of the total DM, Eq.~\eqref{eq:under}.
Additionally, the black lines correspond to different values of $\beta$ required to produce the observed DM abundance via PBH evaporation: $\beta=10^{-10}$ (dotted), $\beta=10^{-15}$ (dash-dotted), and $\beta=10^{-20}$ (solid).
To produce the whole DM density, $10^{-23} \lesssim \beta \lesssim 1$ is required.
It is worth mentioning that such a broad range for $\beta$ can be spanned through a variety of PBH formation mechanisms, e.g., inflationary fluctuations, cosmological phase transitions, or collapse of cosmic loops~\cite{Carr:2020gox}.
For the case of Gaussian density perturbations $\beta\sim \exp[-\delta_c^2/(2\sigma^2)]$ (where $\delta_c\sim\gamma$ is the critical overdensity and $\sigma$ the mass variance at the horizon crossing).
Since usually $\delta_c\gg \sigma$, it follows that small values for $\beta$ are typically expected~\cite{Morrison:2018xla}.
Additionally, the blue dashed line corresponding to $\mdm = \Tbhin$ shows the transition between the two regimes in Eq.~\eqref{eq:Ndm}.
Above that line, DM is light enough to be radiated by PBHs during their whole evaporation; whereas below it, the emission only occurs when PBH temperature is higher than the DM mass.

Before concluding this section, we note that there are regions of the parameter space in Fig.~\ref{fig:contour} where PBHs dominate the energy density of the universe before their full evaporation.
That can be clearly seen by studying the parameter space $[M_\text{in},\,\beta]$.
However, in the plane $[m_\text{DM},\,M_\text{in}]$ used in Fig.~\ref{fig:contour}, it is completely hidden.
Few comments are in order:
$i)$ Above the green dash-dotted lines, the universe is always radiation dominated.
$ii)$ The dash-dotted lines correspond to the case where PBHs eventually dominate.
Here only a lower bound on $\beta$ can be extracted.
$iii)$ Bellow these lines, the value of $\beta$ is not univocally defined as DM can not be fully produced by PBHs.
In that region, the universe could be either PBH- or radiation-dominated.

\section{Gravitational UV freeze-in}
Independently from the PBH evaporation, there is an irreducible DM production channel which is particularly efficient in the region favored by Fig.~\ref{fig:contour}, and corresponds to the gravitational UV freeze-in.
DM can be generated via 2-to-2 annihilations of SM particles, mediated by the exchange of massless gravitons in the $s$-channel.
Its contribution to the total DM density is~\cite{Bernal:2018qlk}
\begin{equation}\label{eq:graviton}
    \frac{\Omega_\text{DM}h^2}{0.12} \lesssim 4.2\times 10^{-13}\,\alpha_\text{DM}\,\frac{\mdm}{1~\text{GeV}}\left(\frac{\Trh}{10^{12}~\text{GeV}}\right)^3,
\end{equation}
where $\alpha_\text{DM}=1.9\times 10^{-4}$, $1.1\times 10^{-3}$ or $2.3\times 10^{-3}$ for scalar, fermionic, or vector DM, respectively.
The equality corresponds to the case where the whole DM abundance is produced via the graviton exchange.
We notice that the DM production has a strong dependence on the reheating temperature, characteristic of the UV freeze-in mechanism.
For simplicity, in Eq.~\eqref{eq:graviton} instantaneous reheating was assumed.
In such approximation, the efficiency of reheating is one~\cite{Garny:2017kha}, and the maximal temperature achieved by the SM plasma is $\Trh$.

We note that the temperatures $\Trh$ and $\Tin$ are in principle unrelated, however, $\Trh \geq \Tin$ to guarantee that PBHs are produced after the onset of the radiation domination era.
Additionally, 2-to-2 scatterings require $\mdm \leq \Tmax$ to be kinematically allowed, where $\Tmax$ corresponds to the maximal temperature reached by the thermal bath~\cite{Giudice:2000ex}.
The relation between $\Trh$ and $\Tmax$ strongly depends on the details of the reheating dynamics.
Here, however, an instantaneous reheating process is assumed, in which case $\Trh=\Tmax$, giving rise to a conservative bound.

In order not to overclose the universe, the loosest upper bound on the DM mass from Eq.~\eqref{eq:graviton} corresponds to $\Tin \simeq \Trh$, whereas the tightest one appears when $\Trh$ takes its highest allowed value $\Trh\simeq 10^{16}$~GeV~\cite{Akrami:2018odb}.
Nevertheless, it is important to note that the DM yield in Eq.~\eqref{eq:graviton} can be significantly boosted when considering a non-instantaneous decay of the inflaton~\cite{Giudice:2000ex, Garcia:2017tuj}, and in particular due to nonthermal effects~\cite{Garcia:2018wtq}, or expansion eras dominated by a fluid component stiffer than radiation~\cite{Bernal:2019mhf, Bernal:2020bfj}.

The lower panel of Fig.~\ref{fig:contour} is equivalent to the upper one, but now considering the irreducible effect of the gravitational freeze-in, taking $\Trh = \Tin$, within the instantaneous decay approximation for the inflaton.
The constraint due to DM overabundance produced by UV freeze-in is overlaid in gray.
We notice that the gravitational UV freeze-in sets strong constraints, excluding large regions of the parameter space favored by PBHs production of DM. 
Here again the black lines correspond to different values of $\beta$ required to produce the observed DM abundance, now via both PBH evaporation and gravitational freeze-in: $\beta=10^{-10}$ (dotted), $\beta=10^{-15}$ (dash-dotted), and $\beta=10^{-20}$ (solid).

\begin{figure}
	\centering
	\includegraphics[height=0.39\textwidth]{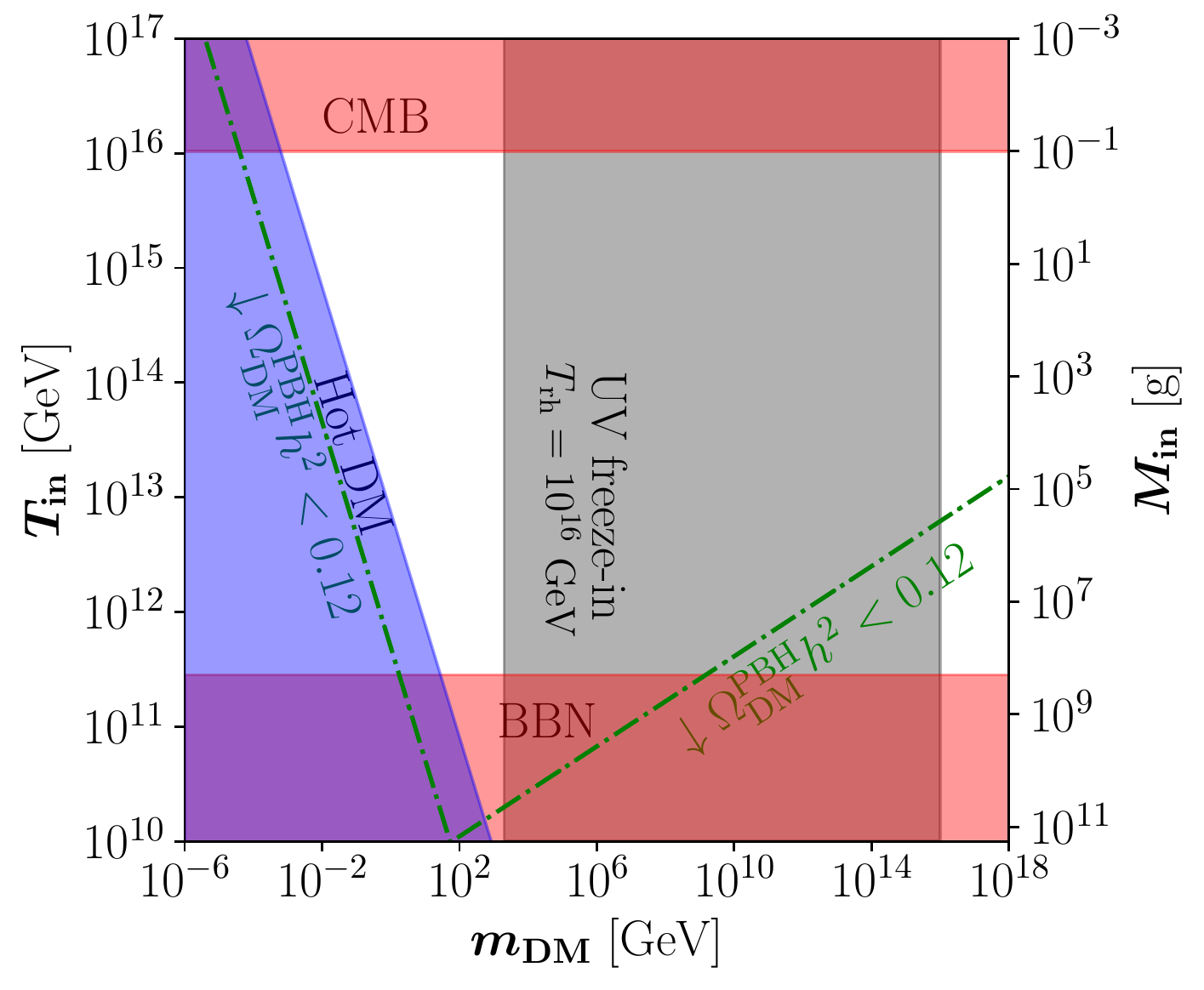}
	\includegraphics[height=0.39\textwidth]{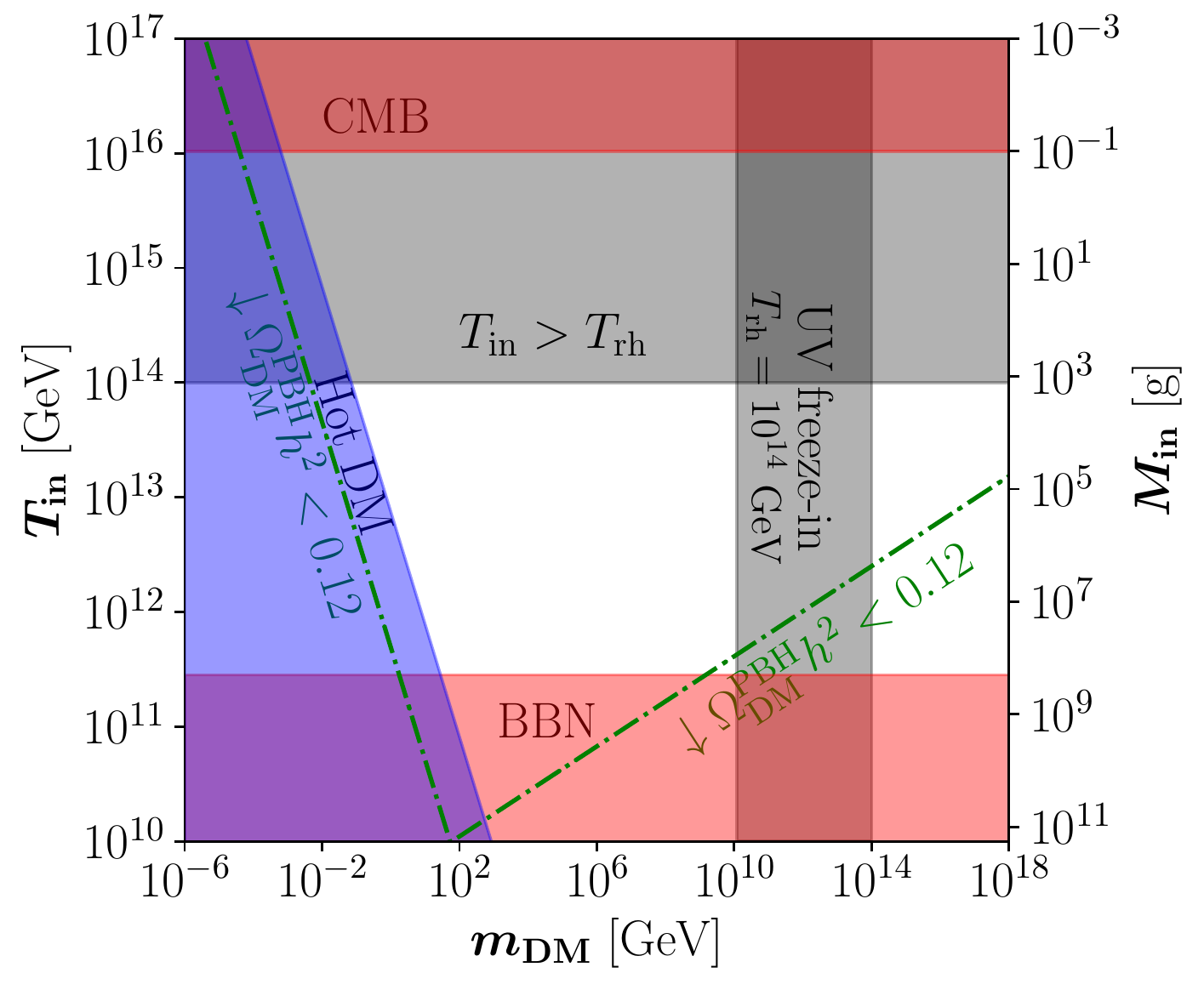}
	\caption{The color shaded areas depict the constraints described in the text.
	Bellow the green dash-dotted lines PBHs produce only a fraction of the total DM.
	In particular, gray areas correspond to the irreducible UV freeze-in DM production via the exchange of gravitons, for $\Trh=10^{16}$~GeV (upper panel), and $\Trh=10^{14}$~GeV (lower panel).}
	\label{fig:DM}
\end{figure}

Figure~\ref{fig:DM} shows in gray the constraint due to DM overabundance produced by UV freeze-in, for $\Trh=10^{16}$~GeV (upper panel), and $\Trh=10^{14}$~GeV (lower panel).
The regions where $\Tin > \Trh$ are also shown.
Before concluding, we notice that in the case of scalar DM, strong model-dependent constraints due to isocurvature perturbations apply, and therefore large regions of the parameter space corresponding to $\mdm \ll H_I$~\cite{Garny:2015sjg, Garny:2017kha} can be already ruled out observationally~\cite{Akrami:2018odb}.

\section{Conclusions}
The observed abundance of dark matter (DM) particles that only interact gravitationally with the standard model (SM) could have been produced in the early universe by Hawking evaporation of primordial black holes (PBH).
This mechanism is viable in a large range of DM masses, spanning up to the Planck scale, and for PBHs produced shortly after the reheating epoch.
However, in this same range of parameters, the irreducible UV gravitational freeze-in production, where DM particles are produced from the SM via the exchange of gravitons, is active and efficient.
We showed that the latter mechanism sets strong bounds, excluding large regions of the parameter space favored by PBH production.
We note that these bounds on super heavy DM could be even stronger if DM couples to the inflaton.
In that case, DM can be efficiently produced by the decay of the inflatons from the heating era, or generated from PBH evaporation.
Additionally, these production mechanisms also lead to strong limits on other allowed interactions between SM and DM particles~\cite{Chianese:2020yjo, Chianese:2020khl}.
Finally, we note that GW induced by early isocurvature fluctuations in the case of an era of PBH domination could be detectable by the present or future generation of GW observatories~\cite{Inomata:2020lmk, Papanikolaou:2020qtd, Domenech:2020ssp}.\footnote{In particular, if the equation-of-state parameter of the Universe undergoes a deep modification at the end of PBH evaporation, scalar (curvature/density) perturbations would source GWs that can be detected by upcoming GW experiments~\cite{Inomata:2020lmk}.}
In particular, the GW spectrum for $\Min \simeq 10^4$~g to $10^8$~g enters the observational window of advanced LIGO~\cite{LIGOScientific:2019vic} and DECIGO~\cite{Seto:2001qf, Yagi:2011wg}, and could be tested in the future~\cite{Domenech:2020ssp}.

Before concluding, we note that if DM features sizable self-interactions, number-changing processes can enhance the PBH DM production while decreasing the mean DM kinetic energy~\cite{Bernal:2020kse}.
In that case, both the `hot DM' and the underabundance ($\Omega_\text{DM}h^2<0.12$) bounds on the left of Fig.~\ref{fig:DM} can be significantly eased.

\vspace{0.5cm}
\acknowledgments
{\bf Acknowledgments.}\\The authors thank Javier Rubio and ``El Journal Club más sabroso'' for fruitful discussions.
NB received funding from Universidad Antonio Nariño grants 2018204, 2019101, and 2019248, the Spanish MINECO under grant FPA2017-84543-P, and the Patrimonio Autónomo - Fondo Nacional de Financiamiento para la Ciencia, la Tecnología y la Innovación Francisco José de Caldas (MinCiencias - Colombia) grant 80740-465-2020.
The work of OZ is supported by Sostenibilidad-UdeA, the UdeA/CODI Grant 2017-16286, and by COLCIENCIAS through the Grant 111577657253.
This project has received funding /support from the European Union's Horizon 2020 research and innovation programme under the Marie Skłodowska-Curie grant agreement No 860881-HIDDeN.

\bibliography{biblio}

\end{document}